\documentclass[11pt]{article}

\usepackage{epsfig,latexsym}
\usepackage{amsmath}

\newcommand{\join}{\text{\textcircled{{\footnotesize 1}}}}
\newcommand{\cojoin}{\text{\textcircled{{\footnotesize 0}}}}

\newcommand{\qed}{\hfill $\Box$}

\newtheorem{theo}{Theorem}
\newtheorem{lemm}{Lemma}
\newtheorem{coro}{Corollary}

\newtheorem{clai}{Claim}[section]
\newtheorem{prop}{Proposition}

\title{Bounded Clique-Width of ($S_{1,2,2}$,Triangle)-Free Graphs}

\def\inst#1{$^{#1}$}

\author{Andreas Brandst\"adt\inst{1} 
\and Suhail Mahfud\inst{2} 
\and Raffaele Mosca\inst{3}
}

\begin{document}

\maketitle

\begin{center}
{\footnotesize
\inst{1} Institut f\"ur Informatik, Universit\"at Rostock, D-18051 Rostock, Germany.\\
\texttt{ab@informatik.uni-rostock.de}

\inst{2} Tishreen University Latakia, Syria.\\

\inst{3} Dipartimento di Economia, Universit\'a degli Studi "G. D'Annunzio", Pescara 65121, Italy.\\
\texttt{r.mosca@unich.it}
}
\end{center}

\begin{abstract}
If a graph has no induced subgraph isomorphic to $H_1$ or $H_2$ then it is said to be ($H_1,H_2$)-free. Dabrowski and Paulusma found 13 open cases for the question whether the clique-width of ($H_1,H_2$)-free graphs is bounded. One of them is the class of ($S_{1,2,2}$,triangle)-free graphs. In this paper we show that these graphs have bounded clique-width. Thus, also ($P_1+2P_2$,triangle)-free graphs have bounded clique-width which solves another open problem of Dabrowski and Paulusma. 
Meanwhile we were informed by Paulusma that in December 2015, Dabrowski, Dross and Paulusma showed that ($S_{1,2,2}$,triangle)-free graphs (and some other graph classes) have bounded clique-width. 
\end{abstract}


\section{Introduction}

The notion of {\em clique-width} of a graph, defined by Courcelle, Engelfriet and Rozenberg (in the context of graph grammars) in \cite{CouEngRos1993}, is a fundamental example of a width parameter on graphs which leads to efficient algorithms for problems expressible in some kind of Monadic Second Order Logic whenever the class of graphs has bounded clique-width \cite{CouMakRot2000}.

The clique-width $cw(G)$ of a graph $G$ is defined as the minimum number of labels needed to construct $G$ by using the following four operations on vertex-labeled graphs:

\begin{itemize}
\item[$(i)$] creating a new vertex $v$ with (integer) label $\ell$ (denoted by $\ell(v)$).
\item[$(ii)$] taking the disjoint union of two (vertex-labeled and vertex-disjoint) graphs $G_1$, $G_2$ (denoted by $G_1 \oplus G_2$).
\item[$(iii)$] adding all edges between the set of all vertices with label $i$ and the set of all vertices
             with label $j$ for $i \neq j$ (denoted by $\eta_{i,j}$).
\item[$(iv)$] renaming label $i$ to $j$ (denoted by $\rho_{i \rightarrow j}$).
\end{itemize}

A {\em $k$-expression} for a graph $G$ of clique-width $k$ describes the recursive generation of $G$ by repeatedly applying these operations $(i)-(iv)$ using at most $k$ pairwise different labels. 

See \cite{KamLozMil2009} for a survey on clique-width.

Step $(iii)$ is also called {\em join} between labels $i$ and $j$, and renaming labels is also called re-labeling.  

For a subset $M \subset V$, a vertex $z \notin M$ {\em distinguishes} $M$ if there are $x,y \in M$ with $xz \in E$ and $yz \notin E$.   
A subset $M \subset V$ is a \emph{module} if for every vertex $v \notin M$, $v$ does not distinguish $M$. A module $M$ is {\em trivial} if either $M=\emptyset$, $M=V$ or $|M|=1$. A nontrivial module is a \emph{homogeneous} set. Obviously, a vertex set $H$ is homogeneous in $G$ if and only if $H$ is homogeneous in the complement graph $\overline{G}$. A graph is {\em prime} if it does not contain any homogeneous set. In particular, if $G$ is a prime graph then $G$ and $\overline{G}$ are connected.

In \cite{CouMakRot2000,CouOla2000}, various fundamental clique-width properties are shown, among them:

\begin{prop}[\cite{CouMakRot2000,CouOla2000}]\label{cwdmaxprim}
For a graph $G$, $cw(G) = \max\{cw(H): H$ is a prime subgraph of~$G\}$.
\end{prop}

Thus we can focus on prime graphs. Moreover, vertex deletion preserves bounded clique-width; more exactly: 

\begin{prop}\label{cwdconstvertexadd}
If ${\cal C}$ is a class of bounded clique-width and ${\cal C}'$ results from adding a constant number of vertices to all graphs in ${\cal C}$ then also ${\cal C}'$ has bounded clique-width.  
\end{prop}

In \cite{DabPau2014}, Dabrowski and Paulusma analyzed the clique-width of $H$-free bipartite graphs for any $H$, and in \cite{DabPau2015}, they analyzed the clique-width of ($H_1,H_2$)-free graphs and found 13 open cases for the question whether the clique-width of ($H_1,H_2$)-free graphs is bounded. 

Let $P_k$ denote the chordless path $P$ with $k$ vertices, say $a_1,\ldots,a_k$, and $k-1$ edges $a_ia_{i+1}$, $1 \le i \le k-1$; we also denote it as $P=(a_1,\ldots,a_k)$. Let $C_k$ denote the chordless cycle with $k$ vertices. $K_3$ (called {\em triangle}) is the complete graph with three vertices. 
 
For indices $i,j,k \ge 0$, let $S_{i,j,k}$ denote the graph with vertices $u,x_1,\ldots,x_i$, $y_1,\ldots,y_j$, $z_1,\ldots,z_k$ such that the subgraph induced by $u,x_1,\ldots,x_i$ forms a $P_{i+1}$ $(u,x_1,\ldots,x_i)$, the subgraph induced by $u,y_1,\ldots,y_j$ forms a $P_{j+1}$ $(u,y_1,\ldots,y_j)$, and the subgraph induced by $u,z_1,\ldots,z_k$ forms a $P_{k+1}$ $(u,z_1,\ldots,z_k)$, and there are no other edges in $S_{i,j,k}$. Thus, {\em claw} is $S_{1,1,1}$, and $P_k$ is isomorphic to e.g. $S_{0,0,k-1}$. $P_1+2P_2$ denotes the disjoint union of one vertex and two $P_2$'s. Note that $P_1+2P_2$ is an induced subgraph of $S_{1,2,2}$.

One of the open cases in \cite{DabPau2015} is the class of ($S_{1,2,2}$,triangle)-free graphs; it is open even for ($P_1+2P_2$,triangle)-free graphs. 
In a similar case, in \cite{BraKleMah2006}, it was shown that the clique-width of ($P_6$,triangle)-free graphs is bounded, and in \cite{Mahfu2005}, it was shown that the clique-width of ($S_{1,1,3}$,triangle)-free graphs is bounded. 

Based on \cite{Mahfu2005} and \cite{BraKleMah2006}, in this paper we show that ($S_{1,2,2}$,triangle)-free graphs have bounded clique-width. Thus, also the open problem for ($P_1+2P_2$,triangle)-free graphs is solved. 

In \cite{Lozin2002}, Lozin showed that the clique-width of bipartite $S_{1,2,3}$-free graphs is at most 5.  
Thus, we consider prime ($S_{1,2,2},K_3$)-free graphs containing an odd cycle. 

\section{($S_{1,2,2}$,$K_3$)-free graphs containing a $C_5$}

Similarly as for ($P_6,K_3$)-free graphs (see \cite{BraKleMah2006}), the structural properties of ($S_{1,2,2}$,$K_3$)-free graphs containing a $C_5$ are the basic ones for showing bounded clique-width. 

\subsection{Structural properties}

Let $C$ be a $C_5$ in $G$ with vertices $v_1,v_2,v_3,v_4,v_5$ and edges $v_iv_{i+1}$ (index arithmetic modulo~5). 
A {\em $k$-vertex} of $C$, $k \in \{0,1,\ldots,5\}$, is a vertex $v \notin V(C)$ having exactly $k$ neighbors in $V(C)$. 
Since $G$ is triangle-free, $C$ has no $k$-vertex for $k \ge 3$, and every 2-vertex of $C$ has non-consecutive neighbors in $C$. Let $N$ denote the set of 0-vertices of $C$, let $I_i$ denote the set of 1-vertices of $C$ being adjacent to $v_i$ and let $I_{i,j}$ denote the set of 2-vertices of $C$ being adjacent to $v_i$ and $v_j$.
Clearly, since $G$ is triangle-free, $I_i$ and $I_{i,j}$ are independent vertex sets. 
 
\begin{lemm}\label{basicC5prop1}
For every $i \in \{1,\ldots,5\}$ we have: 
\begin{enumerate}
\item[$(i)$] $I_i \join I_{i+1}$ and $I_i \cojoin I_{i+2}$. 
\item[$(ii)$] $I_i \join I_{i-1,i+1}$ and $I_i \cojoin I_{i,i+2} \cup I_{i-2,i}$. 
\item[$(iii)$]  $N \cojoin I_i$ and $N \cojoin I_{i,i+2}$ and thus, $N = \emptyset$. 
\item[$(iv)$]  $I_{i,i+2} \cojoin (I_{i+2,i+4} \cup I_{i-2,i})$. 
\end{enumerate}
\end{lemm} 

{\bf Proof.}
$(i)$: Without loss of generality, let $x \in I_1$, $y \in I_2$ and $z \in I_3$. Since $\{x,v_1,v_2,v_3,v_4,y\}$ does not induce an $S_{1,2,2}$, we have $xy \in E$. Since $\{x,z,v_3,v_4,v_5,v_2\}$ does not induce an $S_{1,2,2}$, we have $xz \notin E$. 

$(ii)$: $I_i \join I_{i-1,i+1}$: Without loss of generality, let $x \in I_1$, and $y \in I_{5,2}$. Since $\{x,v_1,v_2,v_3,v_4,y\}$ does not induce an $S_{1,2,2}$, we have $xy \in E$.

$I_i \cojoin I_{i,i+2} \cup I_{i-2,i}$: Holds since $G$ is triangle-free. 

$(iii)$: Let $z \in N$, and without loss of generality, let $x \in I_1$, and $y \in I_{2,4}$. Since $\{z,x,v_1,v_2,v_3,v_5\}$ does not induce an $S_{1,2,2}$, we have $xz \notin E$, and since $\{z,y,v_4,v_5,v_1,v_3\}$ does not induce an $S_{1,2,2}$, we have $yz \notin E$. Thus, since $G$, as a prime graph, is connected, we have  
$N = \emptyset$. 

$(iv)$: Holds since $G$ is triangle-free. 
\qed

\begin{coro}\label{C5Sidisting}
For every $i \in \{1,\ldots,5\}$ we have: 
\begin{enumerate}
\item[$(i)$] $I_i$ can only be distinguished by vertices in $I_{i+1,i+3} \cup I_{i-3,i-1}$. 
\item[$(ii)$] $I_{i,i+2}$ can only be distinguished by vertices in $I_{i+1,i+3} \cup I_{i-1,i+1} \cup I_{i+3} \cup I_{i-1}$. 
\end{enumerate}
\end{coro} 

Let 
\begin{enumerate}
\item[ ] $I'_i:=\{x \in I_i: x \cojoin I_{i+1,i+3} \cup I_{i-3,i-1}\}$. 
\end{enumerate}

Since $G$ is prime, we have $|I'_i| \le 1$ for all $i \in \{1,\ldots,5\}$. Thus, by Proposition \ref{cwdconstvertexadd}, from now on, we can assume that 
$I'_i=\emptyset$: 

\begin{enumerate}
\item[$(A)$] For all $i \in \{1,\ldots,5\}$, every vertex in $I_i$ has a neighbor in $I_{i+1,i+3} \cup I_{i-3,i-1}$. 
\end{enumerate}

Clearly, if $x \in I_{i+1,i+3}$ has a neighbor in $I_i$ then, since $G$ is triangle-free and by Lemma \ref{basicC5prop1} $(i)$, we have $x \cojoin I_{i-1}$.  

Let 
\begin{enumerate}
\item[ ] $I'_{i,i+2}:=\{x \in I_{i,i+2}: x \join I_{i+1,i+3} \cup I_{i-1,i+1}$ and $x \cojoin I_{i-1} \cup I_{i+3}\}$. 
\end{enumerate}

Then obviously, $I'_{i,i+2} \cup \{v_{i+1}\}$ is a module. Since $G$ is prime, we have $I'_{i,i+2}=\emptyset$. Thus, from now on, we can assume that 
\begin{enumerate}
\item[$(B)$] every vertex in $I_{i,i+2}$ has either a non-neighbor in $I_{i+1,i+3} \cup I_{i-1,i+1}$ or a neighbor in $I_{i+3} \cup I_{i-1}$. 
\end{enumerate}

A bipartite graph $B=(X,Y,E)$ is a {\em bipartite chain graph} if for every $x,x' \in X$, either $N_Y(x) \subseteq N_Y(x')$ or $N_Y(x') \subseteq N_Y(x)$. It is well known that $B$ is a bipartite chain graph if and only if $B$ is $2P_2$-free, and the clique-width of bipartite chain graphs is at most 3 (e.g., since bipartite chain graphs are distance hereditary and the clique-width of distance-hereditary graphs is at most 3 - see \cite{GolRot2000}). 

\begin{lemm}\label{basicC5S122prop2}
For every $i \in \{1,\ldots,5\}$ we have: 
\begin{enumerate}
\item[$(i)$] There is no independent triple $x,y,z$ with $x \in I_i$, $y \in I_{i+1,i+3}$, and $z \in I_{i-3,i-1}$. 
\item[$(ii)$] $G[I_i \cup I_{i+1,i+3}]$ $(G[I_i \cup I_{i-3,i-1}]$, respectively$)$ is a bipartite chain graph.
\item[$(iii)$] $G[I_{i,i+2} \cup I_{i+1,i+3}]$ is a bipartite chain graph.
\item[$(iv)$] For each vertex $x \in I_{i,i+2}$ having a non-neighbor in $I_{i+1,i+3}$ $($in $I_{i-1,i+1}$, respectively$)$, we have $x \join I_{i-1,i+1}$ $(x \join I_{i+1,i+3}$, respectively$)$. In particular, there is no independent triple $a \in I_{i,i+2}$, $b \in I_{i+1,i+3}$, and $c \in I_{i+2,i+4}$.
\item[$(v)$] For each vertex $x \in I_{i,i+2}$ having a neighbor in $I_{i+3}$ $($in $I_{i-1}$, respectively$)$, we have $x \join I_{i+1,i+3}$ $(x \join I_{i-1,i+1}$, respectively$)$ and $x \cojoin I_{i-1}$ $(x \cojoin I_{i+3}$, respectively$)$.
\end{enumerate}
\end{lemm} 

{\bf Proof.}
$(i)$: Without loss of generality, let $x,y,z$ be an independent triple with $x \in I_1$, $y \in I_{2,4}$ and $z \in I_{3,5}$. 
Then $\{y,v_2,v_1,v_5,z,x\}$ induces an $S_{1,2,2}$, which is a contradiction. 

$(ii)$: Without loss of generality, let $x,x' \in I_1$, and $y,y' \in I_{2,4}$, and suppose that $xy \in E$, $xy' \notin E$, $x'y \notin E$, and $x'y' \in E$. 
Then $\{x,y,v_4,y',x',v_5\}$ induces an $S_{1,2,2}$, which is a contradiction. 

$(iii)$: Without loss of generality, let $x,x' \in I_{1,3}$, and $y,y' \in I_{2,4}$, and suppose that $xy \in E$, $xy' \notin E$, $x'y \notin E$, and $x'y' \in E$. 
Then $\{x,y,v_4,y',x',v_5\}$ induces an $S_{1,2,2}$, which is a contradiction. 

$(iv)$: Without loss of generality, let $x \in I_{1,3}$ and $y \in I_{2,4}$ with $xy \notin E$. Suppose that there is a vertex $z \in S_{5,2}$ with $xz \notin E$.
Then $\{x,v_3,v_4,v_5,z,y\}$ induces an $S_{1,2,2}$, which is a contradiction. 

$(v)$: Without loss of generality, let $x \in I_{1,3}$ and $y \in I_4$ with $xy \in E$. Suppose that there is a vertex $z \in I_{2,4}$ with $xz \notin E$.
Then $\{x,y,v_4,z,v_2,v_5\}$ induces an $S_{1,2,2}$, which is a contradiction. The second condition holds since $G$ is triangle-free and by Lemma \ref{basicC5prop1}
$(i)$. 
\qed

\subsection{3-chain graphs}

Now, as a first step, we describe a generalization of bipartite chain graphs which is closely related to ($S_{1,2,2},K_3$)-free graphs: 
 
$G=(V,E)$ is a {\em $3$-chain graph} if $V$ has a partition into three independent sets $A,B,C$ such that $|A|=|B|=|C|=p$, say $A=\{a_1,\ldots,a_p\}$, $B=\{b_1,\ldots,b_p\}$, $C=\{c_1,\ldots,c_p\}$, and 
\begin{enumerate}
\item[$(i)$] Every pair of $A,B,C$ induces a bipartite chain graph in $G$.  
\item[$(ii)$] 
\begin{enumerate}
\item For all $i \in \{1,\ldots,p\}$, we have $N_B(a_i)=\{b_1,\ldots,b_i\}$, $N_C(a_i)=\{c_i,\ldots,c_p\}$, and 
\item for all $i \in \{1,\ldots,p-1\}$, we have $N_B(c_i)=\{b_{i+1},\ldots,b_p\}$, and $N_B(c_p)=\emptyset$.
\end{enumerate}
\end{enumerate}

In particular, we have: 
\begin{enumerate}
\item[ ] Every vertex in $A$ has a neighbor in $B$ and a neighbor in $C$. 
\item[ ] Every vertex in $B$ has a neighbor in $A$ and a non-neighbor in $C$. 
\item[ ] Every vertex in $C$ has a neighbor in $A$ and a non-neighbor in $B$. 
\end{enumerate}

\begin{lemm}\label{3chainboundedcwd}
The clique-width of $3$-chain graphs is at most $6$.
\end{lemm} 

{\bf Proof.}
Let $G=(V,E)$ be a 3-chain graph with $V=A \cup B \cup C$, $A=\{a_1,\ldots,a_p\}$, $B=\{b_1,\ldots,b_p\}$, $C=\{c_1,\ldots,c_p\}$, as defined above.
We construct $G$ as follows:

\begin{enumerate}
\item Create $a_1$ with label $l_1$, $b_1$ with label $l_2$, and $c_1$ with label $l_3$. 
\item Join $l_1$ with $l_2$ and $l_1$ with $l_3$. 
\item {\bf For} $i:=2$ {\bf to} $p$ {\bf do begin}
\begin{enumerate}
\item Create $a_i$ with label $l_4$, $b_i$ with label $l_5$, and $c_i$ with label $l_6$.   
\item Join $l_4$ with $l_2,l_5$, and $l_6$. Join $l_1$ with $l_6$ and $l_5$ with $l_3$. 
\item Relabel $l_4$ to $l_1$, $l_5$ to $l_2$, and $l_6$ to $l_3$. 
\end{enumerate}
\item[ ] {\bf end}
\end{enumerate}
\qed

$3$-chain graphs are closely related to ($S_{1,2,2}$,$K_3$)-free graphs for the following reason:

Let 
\begin{enumerate}
\item[ ] $X_i:=\{v \in I_{i+1,i+3}: v$ has a neighbor in $I_i$ or a non-neighbor in $I_{i+2,i+4}\}$ 
\item[ ] $Y_i:=\{v \in I_{i+2,i+4}: v$ has a neighbor in $I_i$ or a non-neighbor in $I_{i+1,i+3}\}$ 
\item[ ] $Z_i:=\{v \in I_i: v$ has a neighbor in $I_{i+1,i+3} \cup I_{i+2,i+4}\}$ 
\end{enumerate}

Let $H_i:=G[X_i \cup Y_i \cup Z_i]$. By Lemmas \ref{basicC5prop1} and \ref{basicC5S122prop2}, $V(G) \setminus V(C)$ can be partitioned into $X_1,\ldots,X_5$, $Y_1,\ldots,Y_5$, $Z_1,\ldots,Z_5$ such that for any $i,j$ with $i \neq j$, $H_i$ and $H_j$ cannot distinguish each other.  

Now we focus on the following typical case of $H_1$: As before, let $(v_1,...,v_5)$ be a $C_5$ in $G$, let $Z=I_1$ denote the set of 1-vertices adjacent to $v_1$, and let $X=I_{2,4}=\{x_1,\ldots,x_k\}$ ($Y=I_{3,5}=\{y_1,\ldots,y_k\}$, respectively) be the set of 2-vertices adjacent to $v_2,v_4$ (to $v_3,v_5$, respectively).

Clearly, by Lemma \ref{basicC5S122prop2} $(iii)$, $X \cup Y$ induce a bipartite chain graph. For each $i$, $1 \le i \le k$, let $N_Y(x_i)=\{y_1,\ldots,y_{i-1}\}$, and in particular, $N_Y(x_1)=\emptyset$.   

\begin{prop}\label{basicC5S122prop3}
If $zx_j \in E$ then $zx_i \in E$ for all $i \le j$.
\end{prop} 

{\bf Proof.}
Assume that $zx_j \in E$. Then, since $G$ is triangle-free, $z \cojoin N_Y(x_j)$. Thus, by Lemma \ref{basicC5S122prop2} $(i)$, $z$ is adjacent to every non-neighbor of any vertex in $N_Y(x_j)$, i.e., $zx_i \in E$ for all $i \le j$.
\qed

Let $i_z$ be the maximum index for which $z$ is adjacent to $x_i$. Thus, $N_X(z)=\{x_1,\ldots,x_{i_z}\}$, $y_{1},\ldots,y_{i_z-1}$ are non-adjacent to $z$ and 
$y_{i_z+1},\ldots,y_k$ are adjacent to $z$.

In a 3-chain graph, $zy_{i_z} \notin E$ while in general, $zy_{i_z} \in E$ is possible. Thus, 3-chain graphs are a special case of induced subgraphs in the prime 
($S_{1,2,2}$,$K_3$)-free graph containing a $C_5$. 

Now we focus on more details. 

\subsection{Further properties when $G$ contains a $C_5$}

\begin{clai}\label{claim1}
If $I_{i-1,i+1} \neq \emptyset$ and $I_{i+1,i+3} \neq \emptyset$, then for each vertex $x \in I_{i,i+2}$ either $x \join I_{i-1,i+1}$ or $x \join I_{i+1,i+3}$.
\end{clai} 

{\em Proof.} It follows by Lemma \ref{basicC5S122prop2} $(iv)$. 
\qed

Then by Claim \ref{claim1}, $I_{i,i+2}$ admits a partition $\{I^-_{i,i+2},I^+_{i,i+2},I^*_{i,i+2}\}$, where:

\begin{itemize}
\item[ ] $I^-_{i,i+2} := \{x \in I_{i,i+2}: x$ has a non-neighbor in $I_{i+1,i+3}\}$,
\item[ ] $I^+_{i,i+2} := \{x \in I_{i,i+2}: x$ has a non-neighbor in $I_{i-1,i+1}\}$,
\item[ ] $I^*_{i,i+2} := I_{i,i+2} \setminus (I^-_{i,i+2} \cup I^+_{i,i+2})$.
\end{itemize}

{\em Remark $1$.} Note that for every  $i \in \{1, \ldots, 5\}$, $I^-_{i,i+2}$ is non-empty only if $I_{i+1,i+3}$ is non-empty, and $I^+_{i,i+2}$ is non-empty only if $I_{i-1,i+1}$ is non-empty. 
\qed

Then by Lemma \ref{basicC5S122prop2} $(ii)$ and $(iv)$, the following properties hold for every  $i \in \{1, \ldots, 5\}$:

On one hand we have:
\begin{itemize}
\item[ ] $I^-_{i,i+2} \join I_{i-1,i+1}$.
\item[ ] $I^+_{i,i+2} \join I^+_{i-1,i+1} \cup I^*_{i-1,i+1}$. 
\item[ ] $I^+_{i,i+2} \cup I^-_{i-1,i+1}$ induce a $2P_2$-free bipartite subgraph. 
\item[ ] $I^*_{i,i+2} \join I_{i-1,i+1}$. 
\end{itemize}

On the other hand, by symmetry, we have:

\begin{itemize}
\item[ ] $I^-_{i,i+2} \join I^-_{i+1,i+3} \cup I^*_{i+1,i+3}$. 
\item[ ] $I^-_{i,i+2} \cup I^+_{i+1,i+3}$ induce a $2P_2$-free bipartite subgraph. 
\item[ ] $I^+_{i,i+2} \join I_{i+1,i+3}$. 
\item[ ] $I^*_{i,i+2} \join I_{i+1,i+3}$. 
\end{itemize}

\begin{clai}\label{claim2}
For every  $i \in \{1, \ldots, 5\}$, we have $I_i \cojoin I^+_{i+1,i+3} \cup I^-_{i-3,i-1}$.
\end{clai} 

{\em Proof.} Assume to the contrary that without loss of generality, there are $x \in I_1$ and $y \in I^+_{2,4}$ with $xy \in E$. Let $z \in I_{1,3}$ be a non-neighbor of $y$. Then $x,y,z,v_3,v_1,v_5$ induce an $S_{1,2,2}$ in $G$ which is a contradiction. Thus, $I_i \cojoin I^+_{i+1,i+3}$. 
By symmetry, we also have $I_i \cojoin I^-_{i-3,i-1}$. 
\qed

\begin{clai}\label{claim3}
For each vertex $x \in I_i$, either $x \cojoin I^*_{i+1,i+3}$ or $x \cojoin I^*_{i-3,i-1}$.
\end{clai} 

{\em Proof.} Without loss of generality, let $x \in I_1$. If neither $x \cojoin I^*_{2,4}$ nor $x \cojoin I^*_{3,5}$ then $x$ together with neighbors $y \in I^*_{2,4}$ and $z \in I^*_{3,5}$ would induce a triangle in $G$ which is a contradiction.
\qed

\begin{clai}\label{claim4}
If a vertex $x \in I_i$ contacts $I^*_{i+1,i+3}$ $($contacts $I^*_{i-3,i-1}$, respectively$)$, then $x \join I^-_{i+1,i+3}$ and $x \cojoin I_{i-3,i-1}$ 
$(x \join I^+_{i-3,i-1}$ and $x \cojoin I_{i+1,i+3}$, respectively$)$.
\end{clai} 

{\em Proof.} Without loss of generality, assume that $x \in I_1$ contacts $I^*_{2,4}$, i.e., there is a vertex $y \in I^*_{2,4}$ with $xy \in E$. 

First, suppose to the contrary that $x$ has a non-neighbor $y' \in I^-_{2,4}$. Then $y'$ has a non-neighbor $z \in I_{3,5}$. Since $G$ is triangle-free and 
$xy \in E$ and $yz \in E$, we have $xz \notin E$ but then $x,y',z$ is an independent triple which is a contradiction to Lemma \ref{basicC5S122prop2} $(i)$. 

Second, since $G$ is triangle-free, if $x$ contacts $y \in I^*_{2,4}$ then $x$ is non-adjacent to every vertex $z \in I_{3,5}$ since $xy \in E$ and $y \join I_{3,5}$.

This shows: If a vertex $x \in I_i$ contacts $I^*_{i+1,i+3}$  then $x \join I^-_{i+1,i+3}$ and $x \cojoin I_{i-3,i-1}$.

By symmetry, we have: If a vertex $x \in I_i$ contacts $I^*_{i-3,i-1}$ then $x \join I^+_{i-3,i-1}$ and $x \cojoin I_{i+1,i+3}$. 
\qed

Let 
\begin{itemize}
\item[ ] $I^{left}_i := \{x \in I_i: x$ contacts $I^*_{i+1,i+3}$, $x \join I^-_{i+1,i+3}$, $x \cojoin I^+_{i+1,i+3}$, and $x \cojoin I_{i-3,i-1}\}$,
\item[ ] $I^{right}_i := \{x \in I_i: x$ contacts $I^*_{i-3,i-1}$, $x \join I^+_{i-3,i-1}$, $x \cojoin I^-_{i-3,i-1}$, and $x \cojoin I_{i+1,i+3}\}$,
\item[ ] $I^{both}_i := \{x \in I_i: x$ contacts $I^-_{i+1,i+3} \cup I^+_{i-3,i-1}$, and $x \cojoin I^+_{i+1,i+3} \cup I^*_{i+1,i+3} \cup I^-_{i-3,i-1} \cup I^*_{i-3,i-1}\}$.
\end{itemize}

Then by (A) and by Claims \ref{claim2}, \ref{claim3}, and \ref{claim4}, for every  $i \in \{1, \ldots, 5\}$, $I_i$ admits a partition into $\{I^{left}_i, I^{right}_i, I^{both}_i\}$.

As mentioned in Lemma \ref{basicC5S122prop2} $(v)$, we have:
\begin{clai}\label{claim5}
No vertex of $I^*_{i,i+2}$ contacts both $I^{left}_{i-1}$ and $I^{right}_{i+3}$.
\end{clai} 

Then $V(G) \setminus C$ can be partitioned into the following families of sets:

\begin{itemize}

\item ${\cal F}_1 := \{I^*_{i,i+2} \cup I^{left}_{i-1} \cup I^{right}_{i+3}: i = 1, \ldots,5\}$,

\item ${\cal F}_2 := \{I^{both}_{i} \cup I^-_{i+1,i+3} \cup I^+_{i-3,i-1}: i = 1, \ldots,5\}$.

\end{itemize}

\begin{lemm}\label{claim6}
Each member of ${\cal F}_1$ induces a graph with bounded clique-width.
\end{lemm} 

{\bf Proof.} Without loss of generality, let us consider the prime subgraph $G[I^*_{1,3} \cup I^{right}_{4} \cup I^{left}_{5}]$; in particular, it has no isolated vertices. Then by Claim \ref{claim5}, $I^*_{1,3}$ admits a partition $\{W, Z\}$, where: $W = \{x \in I^*_{1,3}: x$ contacts $I^{right}_4\}$ and $Z = \{x \in I^*_{1,3}: x$ contacts $I^{left}_{5}\}$. Let us recall that $I^{right}_{4} \join I^{left}_{5}$ by Lemma \ref{basicC5prop1} $(i)$. 

Then a $k$-expression for $G[I^*_{1,3} \cup I^{right}_{4} \cup I^{left}_{5}]$, with $k$ bounded, may be defined as follows:

\begin{itemize}
\item[(1)] construct an expression by a disjoint union of two {\em local} expressions as follows: 
  
the first local expression describes $W \cup I^{right}_{4}$ as follows: 
\begin{itemize}
\item[$(i)$] label the vertices of $W \cup I^{right}_{4}$ by a $t$-expression, with $t$ bounded, since $G[W \cup I^{right}_{4}]$ is a $2P_2$-free bipartite subgraph 
by Lemma \ref{basicC5S122prop2} $(ii)$; 

\item[$(ii)$] {\sc re-label} the labels of $W$ by a label $l_1$, and {\sc re-label} the labels of $I^{right}_{4}$ by a label $l_2$;
\end{itemize}

the second local expression describes $Z \cup I^{left}_{5}$ as follows: 
\begin{itemize}
\item[$(i)$] label the vertices of $Z \cup I^{left}_{5}$ by a $t$-expression, with $t$ bounded, since $G[Z \cup I^{left}_{5}]$ is a $2P_2$-free bipartite subgraph 
by Lemma \ref{basicC5S122prop2} $(ii)$; 
\item[$(ii)$] {\sc re-label} the labels of $Z$ by a label $l_3$, and {\sc re-label} the labels of $I^{left}_{5}$ by a label $l_4$;
\end{itemize}

\item[(2)] {\sc join} $l_2$ with $l_4$.  
\end{itemize}

By symmetry we can do it similarly for every $I^*_{i,i+2} \cup I^{left}_{i-1} \cup I^{right}_{i+3}$, $i \in \{1,\ldots,5\}$.  
\qed

\begin{lemm}\label{claim7}
Each member of ${\cal F}_2$ induces a graph with bounded clique-width.
\end{lemm} 

{\bf Proof.} 
Without loss of generality, let us consider the prime subgraph $G[I^-_{2,4} \cup I^+_{3,5} \cup I^{both}_{1}]$; in particular, it has no isolated vertices.  
Let $A = I^-_{2,4}$, $B = I^+_{3,5}$, $Z = I^{both}_{1}$. 
Furthermore let us assume that $A, B, Z$ are non-empty, since otherwise $G[A \cup B \cup Z]$ is bipartite, i.e., $2P_2$-free bipartite by Lemma \ref{basicC5S122prop2}, and Lemma \ref{claim7} directly follows.  

We first consider $G[A \cup B]$. By Lemma \ref{basicC5S122prop2} $(iii)$, $G[A \cup B]$ is $2P_2$-free bipartite. 
Let $A_0:=\{x \in A: x \cojoin B\}$ and $B_0:=\{x \in B: x \cojoin A\}$. These sets might be empty; we assume without loss of generality that $A_0 \neq \emptyset$ and $B_0 \neq \emptyset$. Then 
\begin{equation}\label{A0B0}
A_0 \cojoin B \mbox{ and } B_0 \cojoin A.
\end{equation} 

Moreover, assume without loss of generality that $A_0 \neq A$ and $B_0 \neq B$ (otherwise we can proceed as in Lemma \ref{claim6}).

Then, as one can easily prove, there is a partition $\{A_0,A_1,\ldots,A_p\}$ of $A$, and a partition $\{B_0,B_1,\ldots,B_p\}$ of $B$, with $A_i \neq \emptyset$ and $B_i \neq \emptyset$ for all $i \in \{1,\ldots,p\}$, $p \ge 1$, such that for all $i \in \{1,\ldots,p\}$ and $h = p + 1 - i$, we have:

\begin{equation}\label{ABchaingr}
A_i \cojoin B_0 \cup \ldots \cup B_{h-1} \mbox{ and } A_i \join B_{h} \cup \ldots \cup B_p.
\end{equation} 
\begin{equation}\label{BAchaingr}
B_i \cojoin A_0 \cup \ldots \cup A_{h-1} \mbox{ and } B_i \join A_{h} \cup \ldots \cup A_p.
\end{equation} 

Now we consider $G[Z]$. By Lemma \ref{basicC5S122prop2} $(ii)$, $G[Z \cup A]$ is $2P_2$-free bipartite. Let

\begin{itemize}
  \item[ ] $Z^* := \{z \in Z: z$ does not contact $A\}$.
  \item[ ] $Z_i := \{z \in Z: z$ contacts $A_i$ and $z \cojoin A_{i+1} \cup \ldots \cup A_p\}$ , for $i \in \{0,1,\ldots,p-1\}$.
  \item[ ] $Z_p := \{z \in Z: z$ contacts $A_p\}$.
\end{itemize}

Let us consider the following exhaustive cases. 

{\em Case $1$.} $Z^* = \emptyset$.

Thus, there is a partition $\{Z_0,Z_1,\ldots,Z_p\}$ of $Z$. We need the following properties: 
\begin{equation}\label{prop1}
\mbox{ For each } i \in \{1,\ldots,p\} \mbox{ we have } Z_i \join A_0 \cup A_1 \cup \ldots \cup A_{i-1}.
\end{equation}

{\em Proof.} Let us just prove that $Z_i \join A_{i-1}$, since the remaining cases can be similarly proved by (\ref{ABchaingr}). Suppose to the contrary that $z \in Z_i$ has a non-neighbor $a_{i-1} \in A_{i-1}$. By (\ref{ABchaingr}), we have: $A_i \join B_h \cup B_{h+1}$, while $A_{i-1} \cojoin B_{h}$ and $A_{i-1} \join B_{h+1}$. In particular, since $G$ is triangle-free, $z$ does not contact $B_h \cup B_{h+1}$. Then $z,a_{i-1}$, and any vertex of $B_{h}$ form an independent triple which is a contradiction to Lemma \ref{basicC5S122prop2} $(i)$. 
\qed  

\begin{equation}\label{prop2}
\mbox{ For each } i \in \{1,\ldots,p\} \mbox{ and } h = p + 1 - i \mbox{ we have } Z_i \cojoin B_h \cup \ldots \cup B_p.
\end{equation}

{\em Proof.} It follows by (\ref{ABchaingr}) and since $G$ is triangle-free. 
\qed  

\begin{equation}\label{prop3}
\mbox{ For each } i \in \{0,1,\ldots,p\} \mbox{ and } h = p + 1 - i \mbox{ we have } Z_i \join B_0 \cup B_1 \cup \ldots \cup B_{h - 2}.
\end{equation}

{\em Proof.} By definition of $Z_i$, we have $Z_i \cojoin A_{i+1}$. By (\ref{A0B0}) and (\ref{ABchaingr}), $A_{i+1} \cojoin B_0 \cup B_1 \cup \ldots \cup B_{h - 2}$. Then by Lemma~\ref{basicC5S122prop2} $(i)$, the assertion follows.  
\qed 

Thus, for $i \in \{0,1,\ldots,p\}$, $Z_i$ admits a partition $\{Z_i^-, Z_i^+\}$ where:
\begin{itemize}
    \item $Z_i^- := \{z \in Z_i : z$ has a non-neighbor in $A_i\}$.
    \item $Z_i^+ := \{z \in Z_i : z \join A_i\}$.
\end{itemize}
\begin{equation}\label{prop4}
\mbox{ For each } i \in \{0,1,\ldots,p\} \mbox{ and } h = p + 1 - i \mbox{ we have } Z_i^- \join B_{h-1}.
\end{equation}

{\em Proof.} By definition of $Z_i^-$, every $z \in Z_i^-$ has a non-neighbor in $A_i$. By (\ref{ABchaingr}), $A_i \cojoin B_{h-1}$. Then by Lemma \ref{basicC5S122prop2} $(i)$, the assertion follows.  
\qed  

By definition of $Z_i^+$, we have: 
\begin{equation}\label{prop5}
\mbox{ For each } i \in \{0,1,\ldots,p\}, \mbox{ } Z^+_i \join A_{i}.
\end{equation}

Then by the above, by (\ref{prop1})-(\ref{prop5}), and by Lemma \ref{basicC5S122prop2} $(ii)$,  the following relations hold:

\begin{itemize}
\item[(R1)] 
\begin{itemize}
\item $Z_0^- \cojoin A_1 \cup \ldots \cup A_p$, 
\item $Z_0^- \join B_0 \cup B_1 \cup \ldots \cup  B_p$, and 
\item $G[Z_0^- \cup A_0]$ is a $2P_2$-free bipartite subgraph.
\end{itemize}

\item[(R2)] 
\begin{itemize}
\item  $Z_0^+ \cojoin A_1 \cup \ldots \cup A_p$,
\item $Z_0^+ \join A_0$, $Z_0^+ \join B_0 \cup B_1 \cup \ldots \cup B_{p-1}$, and 
\item $G[Z_0^+ \cup B_p]$ is a $2P_2$-free bipartite subgraph. 
\end{itemize}

\item[(R3)] For $i \in \{1,\ldots,p\}$ and $h = p + 1 - i$, we have: 
\begin{itemize}
\item $Z_i^- \cojoin A_{i+1} \cup \ldots \cup A_p$, $Z_i^- \cojoin B_h \cup \ldots \cup B_p$,
\item $Z_i^- \join A_0 \cup A_1 \cup \ldots \cup A_{i-1}$, $Z_i^- \join B_0 \cup B_1 \cup \ldots  \cup B_{h - 1}$, and 
\item $G[Z_i^- \cup A_i]$ is a $2P_2$-free bipartite subgraph.
\end{itemize}

\item[(R4)] For $i \in \{1,\ldots,p\}$ and $h = p + 1 - i$, we have: 
\begin{itemize}
\item $Z_i^+ \cojoin A_{i+1} \cup \ldots \cup A_p$, $Z_i^+ \cojoin B_h \cup \ldots \cup B_p$,
\item $Z_i^+ \join A_0 \cup A_1 \cup \ldots \cup A_{i}$,  $Z_i^+ \join B_0 \cup B_1 \cup \ldots \cup B_{h - 2}$, and 
\item $G[Z_i^+ \cup B_{h-1}]$ is a $2P_2$-free bipartite subgraph.
\end{itemize}

\end{itemize}

Then, by (R1)-(R4), a $k$-expression, with bounded $k$, may be defined as follows:

{\em Comment}: For simplicity let us indicate the labels by the following colors: {\em white}, {\em green}, {\em black}, {\em pale white}, {\em pale green}, {\em pale black}, {\em blue}. \qed \\

{\bf Procedure Labeling}

{\bf begin}

{\em \{preliminary phase\}}

:: create $A_0 \cup Z^-_0$ and label its vertices by a $t$-expression with $t$ bounded ($A_0 \cup Z^-_0$ induces a $2P_2$-free bipartite graph);

:: {\sc re-label} the labels of $A_0$ by $white$;

:: create $Z^+_0 \cup B_{p}$ and label its vertices by a $t$-expression with $t$ bounded ($Z^+_0 \cup B_{p}$ induces a $2P_2$-free bipartite graph);

:: {\sc re-label} the labels of $B_p$ by {\em black};

:: {\sc re-label} the labels of $Z^-_0$ by {\em pale green};

:: {\sc join} {\em pale green} with {\em black};

:: {\sc re-label} {\em pale green} by $green$;

:: {\sc re-label} the labels of $Z^+_0$ by {\em pale green};

:: {\sc join} {\em pale green} with {\em white};

:: {\sc re-label} {\em pale green} by $green$;

{\em \{main phase\}}

:: {\bf For} $k := 1$ {\bf to} $p$ {\bf do}:

:::: {\bf begin}


:::: create $A_k \cup Z^-_k$ and label its vertices by a $t$-expression with $t$ bounded ($A_k \cup Z^-_k$ induces a $2P_2$-free bipartite graph);

:::: {\sc re-label} the labels of $A_k$ by {\em pale white};

:::: {\sc join} {\em pale white} with $black$;


:::: create $Z^+_k \cup B_{h-1}$ [where $h = p+1-k$] and label its vertices by a $t$-expression with $t$ bounded ($Z^+_k \cup B_{h-1}$ induces a $2P_2$-free bipartite graph);

:::: {\sc re-label} the labels of $B_{h-1}$ [where $h = p+1-k$] by {\em pale black};

:::: {\sc join} {\em pale black} with {\em green};

:::: {\sc re-label} the labels of $Z^-_k$ by {\em pale green};

:::: {\sc join} {\em pale green} with {\em pale black};

:::: {\sc join} {\em pale green} with {\em white};

:::: {\sc re-label} {\em pale green} by $green$;

:::: {\sc re-label} the labels of $Z^+_k$ by {\em pale green};

:::: {\sc join} {\em pale green} with {\em pale white};

:::: {\sc join} {\em pale green} with {\em white};

:::: {\sc re-label} {\em pale green} by $green$;

:::: {\sc re-label} {\em pale white} by $white$;

:::: {\sc re-label} {\em pale black} by $black$;

:::: {\bf end};

:: {\bf end}. 

{\em Case $2$.} $Z^* \neq \emptyset$.

This case can be treated similarly to Case 1 according to the following properties:

\begin{prop}\label{prop6}
If $A_0 = \emptyset$, then $G[Z^* \cup B_p]$ is a $2P_2$-free bipartite subgraph, and $Z^* \join B_0 \cup B_1 \cup \ldots \cup B_{p-1}$.
\end{prop} 

{\em Proof.} It follows by definition of $Z^*$, by (\ref{ABchaingr})-(\ref{BAchaingr}), and by Lemma \ref{basicC5S122prop2} $(i)$. 
\qed 

\begin{prop}\label{prop7}
If $A_0 \neq \emptyset$, then $Z^* \join B_0 \cup B_1 \cup \ldots \cup B_{p}$.
\end{prop}

{\em Proof.} It follows by definition of $Z^*$, by (\ref{ABchaingr})-(\ref{BAchaingr}), and by Lemma \ref{basicC5S122prop2} $(i)$. 
\qed 

Assume that $A_0 = \emptyset$. Then $Z_0 = \emptyset$ as well. Then one can apply an approach similar to that of Case 1 by the following slight modifications in Procedure Labeling:

i) rewrite the {\em preliminary phase} as follows according to Proposition \ref{prop6}:


:: create $Z^* \cup B_p$ and label its vertices by a $t$-expression with $t$ bounded ($Z^* \cup B_p$ induces a  $2P_2$-free bipartite graph);

:: {\sc re-label} the labels of $Z^*$ by {\em blue};

:: {\sc re-label} the labels of $B_p$ by {\em black}.

ii) add the following lines after the {\em main phase} according to Proposition \ref{prop6}:

:: {\sc join} $blue$ with $black$. \\

Assume that $A_0 \neq \emptyset$. Then one can apply an approach similar to that of Case 1 by the following slight modifications in Procedure Labeling:

j) add the following lines after the {\em main phase} according to Proposition \ref{prop7}:

:: create $Z^*$ and label its vertices by $blue$;

:: {\sc join} $blue$ with $black$. \\

This completes the proof of Lemma \ref{claim7}.  
\qed

\subsection{Bounded clique-width}

Now we are able to show:
\begin{theo}\label{S122K3frwithC5}
$(S_{1,2,2},K_3)$-free graphs containing a $C_5$ have bounded clique-width.
\end{theo}

{\bf Proof.} Since $C$ has only five vertices, we can restrict to $G[V(G) \setminus V(C)]$ by Proposition \ref{cwdconstvertexadd}, i.e., $V(G) \setminus V(C)$ can be partitioned into the families $F_1$ and $F_2$ of vertex subsets as defined above. Let us say that the {\em sides} of a member $H$ of $F_1$ (or of $F_2$) are the sets whose union defines $H$. For example the sides of $I^*_{i,i+2} \cup I^{right}_{i-2} \cup I^{left}_{i-1}$ are $I^*_{i,i+2}, I^{right}_{i-2}$, and $I^{left}_{i-1}$. Then by Lemmas~\ref{basicC5prop1} and~\ref{basicC5S122prop2}, by definitions, and by Claims \ref{claim2} and \ref{claim4}, we have:

\begin{itemize}
\item[$(*)$] For any side $S$ of any member $H$ of $F_1 \cup F_2$ and for any side $T$ of any member $K$ of $F_1 \cup F_2$ with $K \neq H$, either $S \join T$ or $S \cojoin T$.
\end{itemize}

Note that $F_1 \cup F_2$ has 10 members and 30 sides. Then let $L$ be a set of labels, with $|L| = 30$, such that each label of $L$ is associated to a side of a member of $F_1 \cup F_2$. Then a $k$-expression for $G$, with $k$ bounded, may be defined as follows:

\begin{itemize}
\item[(1)] construct an expression by a {\sc disjoint union} of 10 {\em local} expressions; each local expression describes a member $H$ of $F_1 \cup F_2$ as follows: 
\begin{itemize}
\item[$(i)$] label the vertices of $H$ by a $t$-expression, with $t$ bounded, according to Lemmas \ref{claim6} and \ref{claim7}; 
\item[$(ii)$] for each side $S$ of $H$, {\sc re-label} the labels of $S$, by the label of $L$ associated to $S$;
\end{itemize}
\item [(2)] for each pair ($l_1,l_2$) of labels of $L$, possibly {\sc join} $l_1$ with $l_2$, according to $(*)$.
\end{itemize}

This completes the proof of Theorem \ref{S122K3frwithC5}.  
\qed 

\section{($S_{1,2,2},K_3,C_5$)-free graphs containing an odd cycle}

Now first assume that the prime ($S_{1,2,2},K_3$)-free graph $G$ is $C_5$-free but contains a $C_7$, say $C$ with vertex set $\{v_1,\ldots,v_7\}$ and edges $v_iv_{i+1}$ (index arithmetic modulo 7).

Let $Z$ denote the set of 0-vertices of $C$. Let $I_{i,i+2}$ denote the 2-vertices being adjacent to $v_i$ and $v_{i+2}$.

Then we have:
\begin{lemm}\label{triangleC5S122frC7prop}
For every $i \in \{1,\ldots,7\}$ we have: 
\begin{enumerate}
\item[$(i)$] There are no $i$-vertices of $C$ for $i \in \{1,3,4,5,6,7\}$. 
\item[$(ii)$] The only $2$-vertices are the ones in $I_{i,i+2}$. 
\item[$(iii)$] $Z \cojoin I_{i,i+2}$. 
\item[$(iv)$] $Z =\emptyset$. 
\item[$(v)$] $I_{i,i+2} \cojoin I_{i+2,i+4} \cup I_{i+3,i+5}$. 
\item[$(vi)$] $I_{i,i+2} \join I_{i+1,i+3}$. 
\item[$(vii)$] $I_{i,i+2}=\emptyset$. 
\end{enumerate}
\end{lemm} 

{\bf Proof.}
$(i)$ and $(ii)$: Obvious since $G$ is $(S_{1,2,2},K_3,C_5)$-free. 

$(iii)$: If $z \in Z$ sees $x \in I_{1,3}$ then $z$ would be a 1-vertex for the $C_7$ induced by $\{v_1,\ldots,v_7\} \cup \{x\} \setminus \{v_2\}$ which is a contradiction to $(i)$.

$(iv)$: Since $G$, as a prime graph, is connected, $(iii)$ implies $(iv)$.

$(v)$: Obvious since $G$ is $(K_3,C_5)$-free. 

$(vi)$: If there are $x \in I_{1,3}$ and $y \in I_{2,4}$ with $xy \notin E$ then $x,v_3,v_4,v_5,v_6,y$ induce an $S_{1,2,2}$ which is a contradiction.  

$(vii)$: Obviously follows from $(v)$ and $(vi)$. 
\qed

\begin{coro}\label{S122K3C5frC7}
If the prime $(S_{1,2,2},K_3,C_5)$-free graph $G$ contains a $C_7$ then it is isomorphic to $C_7$. 
\end{coro} 

Finally assume that the prime ($S_{1,2,2},K_3$)-free graph $G$ is $(C_5,C_7,\ldots,C_{2k-1})$-free but contains a $C_{2k+1}$, $k \ge 4$.

Now, the following is easy to see (by using similar arguments as in Lemma \ref{triangleC5S122frC7prop}): 
\begin{coro}\label{S122K3Ck-1frCk+1}
If the prime $(S_{1,2,2},K_3)$-free graph $G$ is $C_{2i-1}$-free for every $i \le k$ but contains a $C_{2k+1}$ then it is isomorphic to $C_{2k+1}$. 
\end{coro} 

Thus we finally have:

\begin{theo}\label{mainboundedcwdS122trianglefr}
$(S_{1,2,2},K_3)$-free graphs have bounded clique-width. 
\end{theo} 

In particular, Maximum Weight Independent Set (MWIS) and various other NP-complete problems can be solved in polynomial or even in linear time for $(S_{1,2,2},K_3)$-free graphs \cite{CouMakRot2000}. In \cite{KarMaf2016}, MWIS is solved in polynomial time for the superclass of $(S_{1,2,2}$,bull)-free graphs.

\medskip

{\bf Acknowledgment.} We thank Dani\"el Paulusma for recently informing us that Theorem \ref{mainboundedcwdS122trianglefr} was shown already in \cite{DabDroPau2015}. 

{\bf Open Problem.} Do $(S_{1,2,3},K_3)$-free graphs have bounded clique-width?

\begin{footnotesize}
\renewcommand{\baselinestretch}{0.4}

\end{footnotesize}

\end{document}